\documentclass{acmtrans2f}
\usepackage{amssymb}
\input epsf
\newcommand{\nat}{\mbox{$I\!\!N$}}
\newcommand{\eps}{\mbox{$\varepsilon$}}
\newcommand{\suf}[1]{\mbox{$_{(#1...)}$}}
\newcommand{\leftq}[2]{\mbox{$cont(#1,#2)$}}
\newcommand{\nxt}{\mbox{\put(3,3){\circle{6}}}\hspace{7.5pt}}
\newcommand{\until}{\mbox{$\,\cal U\,$}}
\newcommand{\eventually}{\mbox{$\Diamond$}}
\newcommand{\always}{\mbox{$\Box$}}
\newcommand{\before}{\mbox{$\,\cal B\,$}}

\newcommand{\modls}{\mbox{$\models_{_{_{\!\!\!\!\!\!RL}}}$}}

\newcommand{\inv}{\mbox{$^{-1}$}}
\newtheorem{theorem}{Theorem}[section]
\newtheorem{lemma}[theorem]{Lemma}

\newdef{definition}[theorem]{Definition}
\newdef{note}[theorem]{Note}
\newdef{remark}[theorem]{Remark}

\firstfoot{ACM Transactions in Computational Logic, Vol.\ TBD, No.\
TBD, TBD TBD, Pages
\pages.}
\runningfoot{ACM Transactions in Computational Logic, Vol.\ TBD, No.\
TBD, TBD TBD.}
\title{Checking Properties within Fairness and Behavior
Abstractions}
\author{ULRICH ULTES-NITSCHE \\ University of Southampton
\and PIERRE WOLPER \\ University of Li\`ege}
\begin{abstract}
This paper is motivated by the fact that verifying liveness
properties under a fairness condition is often problematic,
especially when abstraction is used. It shows that using a more
abstract notion than truth under fairness, specifically the concept
of a property being satisfied \emph{within fairness} can lead to
interesting possibilities. Technically, it is first established that
deciding satisfaction within fairness is a PSPACE-complete problem and
it is shown that properties satisfied within fairness can always be
satisfied by some fair implementation. Thereafter, the interaction
between behavior abstraction and satisfaction within fairness is
studied and it is proved that satisfaction of properties within
fairness can be verified on behavior abstractions, if the abstraction
homomorphism is {\em weakly continuation-closed}.
\end{abstract}
\category{D.2.4}{Software Engineering}{Software/Program
Verification}[Model checking]
\category{F.3.1}{Logics and Meanings of Programs}{Specifying and
Verifying and Reasoning about Programs}[Mechanical verification]
\terms{Theory, Verification}

\keywords{Relative liveness properties, satisfaction within
fairness, behavior abstraction, weakly continuation-closed
homomorphisms}

\begin{document}
\begin{bottomstuff}
Author's addresses: Ulrich Ultes-Nitsche, University of Southampton,
Department of Electronics and Computer Science, Southampton, SO17 1BJ,
United Kingdom. E-mail: uun@ecs.soton.ac.uk.

Pierre Wolper, University of Li\`ege, Institute Montefiore, B28 B-4000
Li\`ege Sart Tilman, Belgium. E-mail: pw@montefiore.ulg.ac.be

This article is the full, improved, and extended
version of the extended abstract \emph{Relative Liveness and
Behaviour Abstraction} that received the \emph{best student paper
award} of PODC'97 \cite{nitschewolper97}. Parts of this work were
done while Ulrich Ultes-Nitsche visited the University of Li\`ege
under a DAAD-fellowship HSP II/AUFE. Ulrich Ultes-Nitsche's former
name was Ulrich Nitsche.
\permission{TBD}{TBD}
\end{bottomstuff}
\maketitle

\section{Introduction}

To be able to verify liveness properties of a
system \cite{alpernschneider85}, it is almost always necessary to
include a fairness hypothesis in the system description
\cite{francez86}. Indeed, introducing a fairness hypothesis makes it
possible to ignore behaviors that correspond to extreme execution
scenarios and that, in any case, would not occur in any reasonable
implementation. Even though this intuition is clear, making fairness
precise is somewhat more complicated: should one be ``weakly'' or
``strongly'' fair, ``transition'' or ``process'' fair, or isn't
``justice'' or even ``compassion'' what fairness should really
be \cite{mannapnueli92}? Of course, there is a rational way of
choosing which fairness notion is adequate for a given problem by
considering the nature of the model being used and making
reasonable assumptions about how it might be implemented, but it
remains that this choice is crucial and delicate.

Furthermore, introducing a fairness hypothesis often makes the
verification process somewhat more problematic. This is especially
true when abstraction is used. Indeed, since after moving to the
abstract level one deals with a reduced set of observables, it can
become impossible to express correctly the fairness hypothesis under
which the system is correct. This makes one wish for a more general and
abstract notion of truth under fairness that would contribute to
simplifying verification, especially in the context of abstraction.
Intuitively, the notion to be formalized is that of a property
being true provided one is given ``some control'' over the choices
made during infinite executions. In other words, one wants to
characterize the properties that can be made true by ``some fair
implementation'' of the system.

In this paper, we show that the concept of a property being satisfied
\emph{within fairness} is a suitable abstraction of truth under fairness
that lends itself easily to verification in the context of
abstraction by using the techniques of
\cite{nitscheochsenschlaeger96a,nitschewolper97,ochsenschlaeger94a,ochsenschlaeger95}.
The idea of satisfaction within fairness is to re-interpret the
notion of \emph{relative liveness properties} as a satisfaction
relation. Relative liveness properties are liveness properties within the
universe of behaviors of the system.  Their definition is a relativized
version of the definition of liveness: every prefix of a behavior of
the system can be extended to an infinite behavior that satisfies the
property. This concept and the dual notion of relative safety
property were introduced in~\cite{henzinger92} as a means of
clarifying the shift from liveness to safety when timing constraints
are introduced in a system. It can also be traced to the notion of 
machine-closed property
\cite{abadilamport88,abadilamport90,alurhenzinger95}. 

Here we make a different use of the concept. In fact, we interpret
relative liveness as a satisfaction relation for properties
represented by temporal logic formulas~\cite{emerson90,pnueli77}.
Notice that for a property to be satisfied within fairness
does correspond, in the desired abstract sense, to the property being
satisfied under fairness. Indeed, in crude terms, the system almost satisfies
properties that are satisfied within fairness: it just needs the ``help of some
fairness'' (remember that every prefix of a behavior of the system can
be extended to an infinite behavior that satisfies the
property). Furthermore, we show that for $\omega$-regular systems and
properties, deciding satisfaction within fairness is a
PSPACE-complete problem. This and the fact that, in a reasonable
sense, properties satisfied within fairness can be satisfied by some
fair implementation  are first indications of the usefulness of this
concept for verification.

This usefulness is even more apparent when considering abstraction.
Indeed, satisfaction within fairness enables us to circumvent the fact
that truth under fairness is usually not preserved by abstraction mappings.
Precisely, we consider abstractions defined by language homomorphisms
in the context of systems described by $\omega$-languages. We prove
that whether a property is satisfied within fairness can be
reliably checked on the abstract system, provided that the
homomorphism is \emph{weakly continuation-closed}. Weakly
contiunation-closed homomorphisms were introduced in
\cite{ochsenschlaeger92} (see also \cite{ochsenschlaeger94a})
where they are called \emph{simple} homomorphisms. For homomorphisms,
being weakly continuation-closed essentially means that they are faithful
with respect to the continuation of a word within a language, i.e. the
image of the continuation is the continuation of the image of the word
in the image of the language. We show that weakly continuation-closed
homomorphisms preserve \emph{exactly} properties satisfied within
fairness.

\section{Introductory Examples}
\label{sec-example}

\begin{figure}[htb]
\centerline{\epsfbox{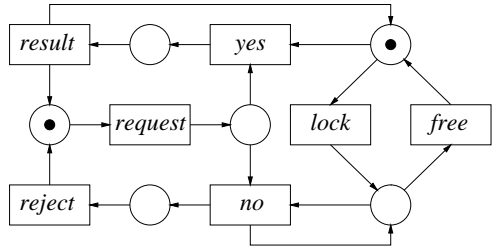}}
\caption{A small system\label{example1}}
\end{figure}

To motivate the definitions we present later on, we start with a small example
of a concurrent reactive system.
Consider the system described as a Petri net in Figure~\ref{example1}.

\begin{figure}[htb]
\centerline{\epsfbox{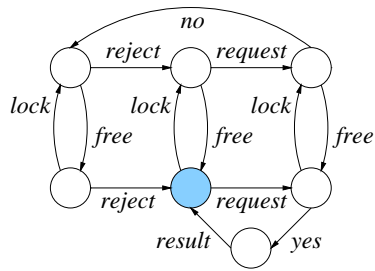}}
\caption{The behaviors of the small system\label{example2}}
\end{figure}

It is a server that, after having received a $request$, can send a
$result$ or a $reject$ion to its client, depending on whether the resource it
manages has been $free$ed 
or
$lock$ed. The possible behaviors of the
system are represented by the finite-state system shown in
Figure~\ref{example2} (the reachability graph of the Petri net). The
initial state is shaded grey, a convention we will also use in subsequent
state diagrams.

From Figure~\ref{example2}, it is easy to see that our system does not
satisfy the propositional linear time temporal
logic~\cite{emerson90,pnueli77} property $\always\eventually(result)$.
Indeed, $lock\cdot(request\cdot no\cdot reject)^\omega$ is a
computation of the system that does not satisfy
$\always\eventually(result)$. Nevertheless, it is clear that what is
missing for the property $\always\eventually(result)$ to be true 
is a fairness hypothesis on the system executions. The
notion of a property being satisfied within fairness captures this:
$\always\eventually(result)$ is satisfied within fairness by the
set of behaviors described by Figure~\ref{example2} (see 
Definition~\ref{uptolive}/\ref{liveof}).

\begin{figure}[htb]
\centerline{\epsfbox{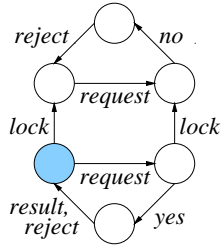}}
\caption{The behaviors of the small system with an error\label{example3}}
\end{figure}

Figure~\ref{example3} gives a finite-state diagram describing the
behaviors of a system similar to the one of Figure~\ref{example1} but
containing an error: in Figure~\ref{example3}, if the resource is
locked, there is no possibility to free it again. There is also
another difference, namely that in Figure~\ref{example3} a request
can also be rejected when the resource is available, but the
motivation for this is linked to our subsequent discussion of
abstraction. The point to notice now, is that no notion of fairness
can make $\always\eventually(result)$ true of the new system and that
the notion satisfaction within fairness captures this again:
$\always\eventually(result)$ is not satisfied within fairness by
the set of behaviors described in Figure~\ref{example3}.

\begin{figure}[htb]
\centerline{\epsfbox{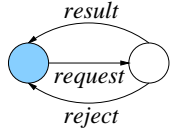}}
\caption{An abstract version of the small system.\label{example4}}
\end{figure}

Let us now consider abstraction. Imagine we are only interested in the
actions $request$, $result$, and $reject$. We thus consider an
abstraction homomorphism that maps all other actions to the empty
word. If we apply this homomorphism to the labeled transition system of
Figure~\ref{example2}, we obtain after reduction the transition diagram
of Figure~\ref{example4}. The property $\always\eventually(result)$ is
satisfied within fairness by the behaviors described in
Figure~\ref{example4}. 

Can we conclude from there that it is also satisfied within fairness
by the behaviors described by Figure~\ref{example2}? Not
without caution since Figure~\ref{example4} is also obtained by
abstracting from Figure~\ref{example3}. What distinguishes the two
abstractions is the nature of the homomorphism. In the case of
Figure~\ref{example2} the homomorphism preserves properties satisfied
within fairness, whereas it does not do so in the case of
Figure~\ref{example3}. In Section~\ref{sec-abstract} we will elaborate
on this and show that 
one can conclude that properties satisfied within
fairness by the abstract system also hold on the concrete
system, precisely when  the homomorphism is weakly continuation-closed,

\section{Preliminaries}\label{sec-prelim}

For defining our concepts, we need several notions from language theory 
\cite{berstel79,eilenberg74,harrison78,thomas90}.
Let $L\subseteq\Sigma^*$ be a language and let $L_\omega\subseteq\Sigma^\omega$
be an $\omega$-language.

\begin{definition}
The \em left quotient \em of $L$ by a word $w\in\Sigma^*$ is defined by
$\leftq{w}{L}=\{v\in\Sigma^*\mid wv\in L\}$.
The left quotient of $L_\omega$ by $w\in\Sigma^*$ is similarly defined
by $\leftq{w}{L_\omega}=\{x\in\Sigma^\omega\mid wx\in L_\omega\}$. 
\end{definition}

The left quotient describes the possible 
continuations of a word in a language. When considering system
behaviors, it describes ``what can happen after $w$ has
happened''. Therefore we denote the left quotient of $L$ by $w$ by
$\leftq{w}{L}$, ``the set of continuations of $w$ in $L$'', instead of
the notation $w\inv(L)$ common in language theory. 

The notation $pre(L)$ designates the set of prefixes of words in $L$.
A language $L$ is called \em prefix-closed \em if and only if
$L=pre(L)$.  For an $\omega$-word $x$, $pre(x)$ designates the set of
all finite prefixes of $x$ and, for
an $\omega$-language 
$L_{\omega}$,
$pre(L_\omega)$ designates the set of all finite prefixes of
$\omega$-words in $L_\omega$.  The Eilenberg-limit \cite{eilenberg74}
of a language $L$ is the set
$lim(L)=\{x\in\Sigma^\omega\mid\exists^\infty w\in pre(x): w\in
L\}$. Here, ``$\exists^\infty...$'' abbreviates: ``there exist
infinitely many different ...''. For a word $w$ and an $\omega$-word
$x$, we denote their $n$th letter by $w_n$ and $x_n$
respectively. Finally, the
notation $x\suf{n}$, $n\in\nat$, 
represents the suffix $x_nx_{n+1}x_{n+2}\ldots$ of an $\omega$-word
$x\in\Sigma^\omega$ starting with the $n^{th}$ letter of $x$.

To describe properties, we use \em propositional linear-time temporal
logic \em (PLTL) \cite{emerson90,pnueli77}.
PLTL-formulas are defined with respect to a set $AP$ of atomic
propositions. All atomic propositions and the 
proposition $true$ are PLTL-formulas. If $\xi$ and $\zeta$ are 
PLTL-formulas, then so are $\neg(\xi)$, $(\xi)\wedge(\zeta)$, $\nxt(\xi)$ 
and $(\xi)\until(\zeta)$.
There exist additional operators that are abbreviations of particular operator 
combinations:
%
%
%
%
%
%
\begin{quotation}
\noindent $(\xi)\vee(\zeta)\equiv\neg((\neg(\xi))\wedge(\neg(\zeta)))$, 

\noindent $(\xi)\Rightarrow(\zeta)\equiv(\neg(\xi))\vee(\zeta)$, 

\noindent $(\xi)\Leftrightarrow(\zeta)\equiv((\xi)\Rightarrow(\zeta))\wedge((\zeta)
\Rightarrow(\xi))$, 

\noindent $\eventually(\xi)\equiv(true)\until(\xi)$, 

\noindent $\always(\xi)\equiv\neg(\eventually(\neg(\xi)))$, 

\noindent $(\xi)\before(\zeta)\equiv\neg((\neg(\xi))\until(\zeta))$.

\end{quotation}

PLTL-formulas are interpreted over infinite sequences of truth values
for the atomic propositions, i.e.\ 
over functions of the type  $\nat \rightarrow
2^{AP}$ or, equivalently over $\omega$-words defined on the alphabet
$2^{AP}$. For convenience, we will also interpret PLTL formulas over
infinite words defined on an arbitrary alphabet $\Sigma$ with the
help of a labeling function $\lambda:\Sigma\rightarrow 2^{AP}$. The
semantics of a PLTL formula with respect to an infinite word $x \in
\Sigma^\omega$ and a labeling function  $\lambda:\Sigma\rightarrow
2^{AP}$ is then the following. (Read ``$\models$'' as ``satisfies.'')
\begin{quotation}
\noindent $x,\lambda \models true$.

\noindent If $\eta$ is an atomic proposition, then $x,\lambda \models\eta$ 
if and only if $\eta\in\lambda(x_{1})$.

\noindent If $\eta=\neg(\xi)$, then $x,\lambda \models\eta$ if and only 
if it is not the case that $x,\lambda \models\xi$.

\noindent If $\eta=(\xi)\wedge(\zeta)$, then $x,\lambda \models\eta$ if and only if $x,\lambda \models\xi$ 
and $x,\lambda \models\zeta$.

\noindent If $\eta=\nxt(\xi)$, then $x,\lambda \models\eta$ if and
only if $x_{(2...)},\lambda \models
\xi$.

\noindent If $\eta=(\xi)\until(\zeta)$, then $x,\lambda \models\eta$ if and only if 
there exists $i\in\nat$ such that $x_{(i...)},\lambda \models\zeta$ and, for 
all $j<i$, $x_{(j...)},\lambda\models\xi$.
\end{quotation}
The meaning of the other operators can be derived from their
definition.  We will write $L_\omega,\lambda\models\eta$ if and only
if $x,\lambda \models\eta$, for all $x\in L_\omega$.

\begin{definition}
A property $\cal P$ over an alphabet $\Sigma$ is a subset of $\Sigma^\omega$. An
$\omega$-language $L_\omega\subseteq\Sigma^\omega$ satisfies $\cal P$
if and only if $L_\omega\subseteq\cal P$. For an alphabet $\Sigma$ and
a labeling function $\lambda: \Sigma \rightarrow 2^{AP}$, the property
represented by a PLTL-formula $\eta$ over $AP$ is the set
$L_\eta=\{x\in\Sigma^\omega\mid x,\lambda \models\eta\}$. 
\end{definition}

\section{Relative Liveness and Safety}

In this section, we review the definition of relative liveness
properties of an $\omega$-language, as well as their counterpart
relative safety properties.  Based on the notion of a relative
liveness property, we will define the satisfaction of properties within
fairness. Let $L_\omega\subseteq\Sigma^\omega$ be
an $\omega$-language representing the behavior of a system and let
${\cal P}\subseteq\Sigma^\omega$ be a property.

\begin{definition}
\label{liveof}
A property ${\cal P}$ is a \em relative liveness property \em
of $L_\omega$ (we write this already as a satisfaction relation:
$L_\omega\modls\cal P$) if and only if\ \ $\forall w\in
pre(L_\omega): \exists x\in \leftq{w}{L_\omega}: wx\in\cal P$.
\end{definition}

\begin{definition}
A property ${\cal P}$ is a \em relative safety property \em of
$L_\omega$ if and only if\ \ $\forall x\in L_\omega$, if $x\not\in
{\cal P}$, then $\exists w\in pre(x): \forall
z\in\leftq{w}{L_\omega}: wz\not\in{\cal P}$.
\end{definition}

\begin{remark}
If $L_\omega=\Sigma^\omega$, then the definitions of relative liveness
and relative safety become exactly the definitions of liveness and
safety given in \cite{alpernschneider85}.
\end{remark}

To prove the decidability of relative liveness and safety 
for regular $\omega$-languages, we use the following
characterizations of these properties.

\begin{lemma}\label{live}
$\cal P$ is a relative liveness property of $L_\omega$ if and only if 
\[pre(L_\omega)=pre(L_\omega\cap{\cal P}).\]
\end{lemma}

\begin{proof}
By definition, $L_\omega\modls{\cal P}$ if 
and only if, for all $w\in pre(L_\omega)$, there exists $x\in 
\leftq{w}{L_\omega}$ such that $wx\in\cal P$. Hence we have $w\in 
pre(L_\omega\cap {\cal P})$, for all $w\in pre(L_\omega)$. This is
equivalent to $pre(L_\omega)\subseteq pre(L_\omega\cap{\cal P})$.
On the other hand,
$pre(L_\omega\cap{\cal P})\subseteq pre(L_\omega)$,
and thus $pre(L_\omega)=pre(L_\omega\cap{\cal P})$.

If $pre(L_\omega)=pre(L_\omega\cap{\cal P})$, then $w\in pre(L_\omega\cap{\cal 
P})$, for all $w\in pre(L_\omega)$.
Therefore,
for all $w\in pre(L_\omega)$, there exists an $x\in
\leftq{w}{L_\omega}$ such that $wx\in\cal P$ and hence $\cal P$ is a
relative liveness property of $L_\omega$.
\qed
\end{proof}

\begin{lemma}\label{safe}
$\cal P$ is a relative safety property of $L_\omega$ if and only if 
\[L_\omega\cap lim(pre(L_\omega\cap{\cal P}))\subseteq\cal P.\]
\end{lemma}

\begin{proof}
By definition, ${\cal P}$ is a relative safety property of $L_\omega$ if 
and only if
\[\forall x\in L_\omega: (\ x\not\in {\cal P} \Rightarrow (\;\exists w\in pre(x): \forall z\in\leftq{w}{L_\omega}: wz\not\in{\cal P}\;)\ ).\]

By taking the counterpositive of the implication this is equivalent to
\[\forall x\in L_\omega: (\ (\;\forall w\in pre(x): \exists z\in\leftq{w}{L_\omega}:
wz\in{\cal P}\;)\Rightarrow x\in {\cal P}\ ).\]

The part
$(\;\forall w\in pre(x): \exists z\in\leftq{w}{L_\omega}:
wz\in{\cal P}\;)$ is equivalent to the condition $pre(x)
\subseteq pre(L_\omega\cap{\cal P})$. Thus, $\cal P$ is a relative
safety property of $L_\omega$ if and only if
$\forall x\in L_\omega: (\ (\;pre(x)\subseteq pre(L_\omega\cap{\cal P})\;)
\Rightarrow x\in {\cal P}\ )$.
All $\omega$-words $x$ in $L_\omega$ such that 
$pre(x)\subseteq pre(L_\omega\cap{\cal P})$ can be represented by 
the set $L_\omega\cap lim(pre(L_\omega\cap{\cal P}))$. Thus, $\cal P$
is a relative safety property of $L_\omega$ if and only if 
$L_\omega\cap lim(pre(L_\omega\cap{\cal P}))\subseteq\cal P$.
\qed
\end{proof}

\begin{theorem}
Given an $\omega$-regular language $L_\omega$ and an
$\omega$-regular property ${\cal P}$ given by nondeterministic B\"uchi
automata or PLTL formulas, determining if ${\cal P}$ is a relative
liveness or safety property is decidable and is a PSPACE-complete
problem.
\end{theorem}

\begin{proof}
The characterizations given by Lemma~\ref{live} and Lemma~\ref{safe}
reduce the problem to questions decidable in
PSPACE~\cite{thomas90,GJ79} (notice that for PLTL formulas one can
build in PSPACE an automaton for the formula and for its
complement~\cite{VW94}). Hardness can be established by a reduction
from 
regular language inclusion~\cite{GJ79}.
\qed
\end{proof}

Note that Lemma~\ref{live} provides the link between relative
liveness and machine closure. Indeed, recall the following
definition~\cite{abadilamport88,abadilamport90,alurhenzinger95}. 

\begin{definition}
Let $\Lambda\subseteq L_\omega\subseteq\Sigma^\omega$, for an alphabet
$\Sigma$. $(L_\omega,\Lambda)$ is called a machine closed live
structure if and only if $pre(L_\omega)\subseteq pre(\Lambda)$.
\end{definition}

We thus have that
$P\subseteq\Sigma^\omega$ is a relative liveness property of
$L_\omega$ if and only if $(L_\omega,P\cap L_\omega)$ is a machine
closed live structure (see Lemma~\ref{live}).

General properties can always be represented as the intersection of a
liveness and a safety property~\cite{alpernschneider85}. As given
precisely below, the relativized version of this result is that a
property holds for an $\omega$-language if it is both a relative
liveness and a relative safety property of the language.

\begin{theorem}
An $\omega$-language $L_{\omega}$ satisfies a property ${\cal P}$
($L_\omega\subseteq{\cal P}$) if and only if ${\cal P}$ is a relative safety
and a relative liveness property of $L_\omega$.
\end{theorem}

\begin{proof}
If $L_\omega\subseteq\cal P$, then, trivially, $\cal P$ is a relative
safety and a relative liveness property of $L_\omega$.

If $\cal P$ is a relative safety property of $L_\omega$, then
$L_\omega\cap lim(pre(L_\omega\cap{\cal P}))\subseteq\cal P$ (Lemma~\ref{safe}).
If, additionally, $\cal P$ is a relative liveness property of
$L_\omega$, then, by  Lemma~\ref{live},
$pre(L_\omega)=pre(L_\omega\cap{\cal P})$. Therefore, we can replace
$pre(L_\omega\cap{\cal P})$ by $pre(L_\omega)$ in the safety condition
and obtain $L_\omega\cap lim(pre(L_\omega))\subseteq\cal P$.
Because $L_\omega\cap lim(pre(L_\omega))=L_\omega$, we finally obtain
$L_\omega\subseteq\cal P$.
\qed
\end{proof}

As shown in  \cite{henzinger92}, 
relative 
liveness and safety
properties also have an elegant definition within the Cantor topology,
i.e. the topological space over $\Sigma^\omega$ compatible with the
following metric \cite{eilenberg74}. (For topological notions see
\cite{kelley55}.)

\begin{definition}
Let $common(x,y)$ designate the longest common prefix of 
two $\omega$-words $x$ and $y$ in $\Sigma^\omega$. We define the 
metric $d(x,y)$ by 
\[\forall x,y\in\Sigma^\omega, x\neq y: d(x,y)=\frac{1}{|common(x,y)|+1}\]
\[\forall x\in\Sigma^\omega: d(x,x)=0.\]
\end{definition}

\begin{lemma}\label{dense}
A property $\cal P$ is a relative liveness property of an
$\omega$-language $L_\omega$ if and only if 
$L_\omega\cap\cal P$ is a dense set in $L_\omega$.
\end{lemma}

\begin{proof}
Let $L_\omega\modls\cal P$, and let $x\in L_\omega$. Then
$pre(L_\omega)=pre(L_\omega\cap{\cal P})$. 
Thus, $pre(x)\subseteq pre(L_\omega\cap {\cal P})$, and we 
have $\forall w\in pre(x): \exists y\in L_\omega\cap{\cal P}: w\in pre(y)$.
We get, for all $x\in L_\omega$ and 
all $\eps>0$ ($\eps$ is related to $\frac{1}{|w|+1}$), that there is a
$y\in L_\omega\cap{\cal P}$ such that $d(x,y)<\eps$. 
So $L_\omega\cap\cal P$ is a dense set in $L_\omega$.

Let $L_\omega\cap\cal P$ be a dense set in $L_\omega$. Then, for all
$x\in L_\omega$ and  all $\eps>0$, there exists $y\in
L_\omega\cap{\cal P}$ such  
that $d(x,y)<\eps$.
Let $x$ be in $L_\omega$, let $w$ be in $pre(x)$ and let 
$\eps=\frac{1}{|w|+1}$.
Because $L_\omega\cap\cal P$ is a dense set in $L_\omega$, there
exists $y\in L_\omega\cap\cal P$ 
such that $w\in pre(y)$. Thus  $pre(L_\omega)\subseteq
pre(L_\omega\cap{\cal P})$. 
Because $pre(L_\omega\cap{\cal P})\subseteq pre(L_\omega)$, we 
have $pre(L_\omega)=pre(L_\omega\cap{\cal P})$. By Lemma~\ref{live}, $\cal P$ is a
relative liveness property of $L_\omega$.
\qed
\end{proof}

\begin{lemma}\label{closed}
A property $\cal P$ is a relative safety property of an
$\omega$-language $L_\omega$ if and only if 
$L_\omega\cap\cal P$ is a closed set in $L_\omega$.
\end{lemma}

\begin{proof}
$\cal P$ is a relative safety property of $L_\omega$ if and only if
\[\forall x\in L_\omega: (\ x\not\in {\cal P} \Rightarrow (\;\exists w\in pre(x): \forall z\in\leftq{w}{L_\omega}: wz\not\in{\cal P}\;)\ ).\]

If $\overline{\cal P}$ is the complement of $\cal P$ with respect 
to $L_\omega$, i.e.\ $\overline{\cal P}=L_\omega\cap(\Sigma^\omega\setminus{\cal P})$, which is 
equivalent 
to $\overline{\cal P}=L_\omega\setminus(L_\omega\cap{\cal P})$, then
$\cal P$ is a relative safety
property of $L_\omega$ if and only if
$\forall x\in L_\omega: (x\in \overline{\cal P} \Rightarrow (\exists w\in pre(x): \forall z\in\leftq{w}{L_\omega}: wz\in\overline{\cal P}))$.
If we define this condition topologically, then $\cal P$ is a relative safety 
property of $L_\omega$ if and only if
$\forall x\in\overline{\cal P}:\exists\eps>0:\forall y\in L_\omega: d(x,y)<\eps
\Rightarrow y\in\overline{\cal P}$.
Thus, $\cal P$ is a relative safety property of $L_\omega$ if and only 
if $\overline{\cal P}$ is an open set in $L_\omega$. 
Because $\overline{\cal P}=L_\omega\setminus(L_\omega\cap{\cal P})$ is the complement 
of $L_\omega\cap\cal P$ with respect to $L_\omega$, we finally 
obtain that $\cal P$ is a relative safety property of $L_\omega$ if and only 
if $L_\omega\cap{\cal P}$ is a closed set in $L_\omega$. 
\qed
\end{proof}

Relative safety having been introduced to complete the picture
around relative liveness, we will now use relative liveness as a
satisfaction relation, calling it satisfaction \emph{within fairness}.

\begin{definition}
\label{uptolive}
We say that $L_\omega$ satisfies $\cal P$ \emph{within fairness} if and
only if $L_\omega\modls\cal P$.
\end{definition}

We have chosen the phrase ``within fairness'' to stress the fact that
for a property satisfied ``within fairness'' to be fully satisfied,
the only missing element is a form of fairness condition on the set
of behaviors being considered. Note that since a safety property
never requires a fairness condition, a safety property satisfied
within fairness by a set of behaviors is also fully satisfied by that
set of behaviors. To prove this, recall the definition of a safety
property (\cite{alpernschneider85}, adapted to our notation):

\begin{definition}
Property ${\cal P}\subseteq\Sigma^\omega$ is called a safety property if
and only if, for all $x\in\Sigma^\omega$, $x\not\models\cal P$ implies
$\exists w\in pre(x):\forall y\in\Sigma^\omega: wy\not\models\cal P$.
\end{definition}

We then have the following.
\begin{lemma}
If $\cal P$ is a safety property, then $L_\omega\modls\cal P$ if and
only if $L_\omega\models\cal P$.
\end{lemma}

\begin{proof}
Let $L_\omega\modls\cal P$, i.e.\ $pre(L_\omega)=pre(L_\omega\cap
{\cal P})$. Assume $L_\omega\not\models\cal
P$. Let $x\in L_\omega$ such that $x\not\models\cal P$. Because $\cal
P$ is a safety property, there exists $w\in pre(x)$ such that $\forall
y\in\Sigma^\omega: wy\not\models\cal P$. So $w$ is not a prefix of an
$\omega$-word in $\cal P$ and thus it is not in $pre(L_\omega\cap{\cal
P})$. Since $w$ is in $pre(L_\omega)$ we have that $pre(L_\omega)\neq
pre(L_\omega\cap {\cal P})$ which contradicts $L_\omega\modls\cal
P$. So $L_\omega\models\cal P$ must hold.

If $L_\omega\models\cal P$, then $L_\omega\modls\cal P$ follows
immediately. \qed
\end{proof}
\section{Implementing Systems that Satisfy Properties within fairness}

If a property is satisfied by a set of behaviors within fairness,
our expectation is that a fair implementation of this set of behaviors
will satisfy the property in the classical sense. Unfortunately, this is
not true for every implementation, even if one assumes strong fairness.
As an example, consider the set of behaviors $\{a,b\}^\omega$. It is
not sufficient to impose strong fairness on the minimal automaton
representing  $\{a,b\}^\omega$ in order to satisfy all properties that
are satisfied within fairness by $\{a,b\}^\omega$. For instance,
$\eventually (a \wedge (\nxt a))$ would not be satisfied, even though
it is satisfied within fairness by $\{a,b\}^\omega$. The reason for
this is that, even if fairness is used, more state information needs
to be kept in order to be able to satisfy the property $\eventually (a
\wedge (\nxt a))$. However, it is always possible to add sufficient state
information to a system in order to turn properties that are satisfied
within fairness into properties that are satisfied in the 
classical sense under fairness.
The following theorem makes this precise.

\begin{theorem}
Let $L_{\omega}$ be a limit closed finite-state set of behaviors (one
accepted by a finite state automaton without acceptance conditions,
i.e.\ by a finite-state labeled transition system)
and let ${\cal P}$ be an $\omega$-regular property. Then, if ${\cal P}$
is satisfied within fairness by $L_{\omega}$, 
there exists a finite-state labeled transition system $\cal A$ such
that the
$\omega$-language accepted by $\cal A$ is $L_\omega$ and all strongly
fair computations in $\cal A$ satisfy ${\cal P}$.
\end{theorem}

\begin{proof}
Since ${\cal P}$ is satisfied by $L_{\omega}$ within fairness,
by Lemma~\ref{live} we have that  $pre(L_\omega)=pre(L_\omega\cap
{\cal P})$. Furthermore, since $L_{\omega}$ is limit closed we have
that   $L_\omega=lim(pre(L_\omega))$ and hence 
\begin{equation}
\label{live-property}
L_\omega=lim(pre(L_\omega\cap {\cal P})). 
\end{equation}
Consider thus a reduced
B\"uchi automaton $A$ accepting $L_\omega\cap {\cal P}$ (by reduced we
mean that states from which no $\omega$-word can be accepted have been
eliminated). The finite-state labeled transition system ${\cal A}$ we
are trying to 
construct is $A$ with its acceptance condition removed. Indeed, by
equation (\ref{live-property}) ${\cal A}$ accepts $L_{\omega}$.
Furthermore, all strongly fair infinite computations of ${\cal A}$
will go infinitely often through a former accepting state of $A$ and thus
will satisfy ${\cal P}$. 
\qed
\end{proof}

The theorem we have just proved gives an interesting insight into
properties satisfied within fairness. They are the properties that fairness
makes true of the system, but possibly at the cost of adding state
information to the system implementation in a noninterfering way, i.e.\ 
without altering the set of limit-closed behaviors of the system.

\section{Behavior Abstractions}

We now turn to the problem of verifying a system using
abstraction. We consider finite-state labeled transition systems
(i.e.\ without
acceptance conditions). Hence the finite-word languages accepted by
the systems we consider are the prefix-closed regular languages, and
the $\omega$-languages they accept are the Eilenberg-limits of
prefix-closed regular languages.

We consider abstractions that hide or rename the actions of our
systems. Precisely, we consider \em abstraction
homomorphisms\em\ that are extensions of alphabetic language
homomorphisms to mappings on finite and infinite words as defined below.

\begin{definition}
Let $h:\Sigma\rightarrow(\Sigma'\cup\{\eps\})$ be a total function
($\eps$ designates the empty word) 
and let $\Sigma^\infty=\Sigma^*\cup\Sigma^\omega$. Then, 
the \em abstraction homomorphism\em\  generated by $h$ is the extension of  $h$
to a mapping $h:\Sigma^\infty\rightarrow\Sigma'^\infty$ defined as
follows. 
For all words $w=w_1w_2w_3\ldots w_n\in\Sigma^*$, $n\in\nat$, 
we define $h(w)=h(w_1)h(w_2)h(w_3)\ldots h(w_n)$.  For 
all $\omega$-words $x=x_1x_2x_3\ldots\in\Sigma^\omega$, we define
$h(x)=h(x_1)h(x_2)h(x_3)\ldots$, if $lim(h(pre(x)))\neq\emptyset$. 
Otherwise, if $lim(h(pre(x)))=\emptyset$, then $h(x)$ is undefined.
\end{definition}

This leads naturally to the following definition of the abstract
behavior of a system under an abstraction homomorphism.

\begin{definition}
Let ${\cal S}$ be a system whose behaviors are the limit $lim(L)$
of a prefix-closed regular language $L$. Then, the \em abstract
behavior\em\  of ${\cal S}$ with respect to the abstraction
homomorphism $h$ is $lim(h(L))$.
\end{definition}

Our goal is to prove properties of the behaviors $lim(L)$ of a
system ${\cal S}$ by only considering the abstract behaviors
$lim(h(L))$ for some abstraction homomorphisms $h$. More specifically,
we are interested in the preservation of properties satisfied within
fairness under the abstraction homomorphism.

Essential information about the properties that are satisfied within
fairness by $lim(L)$ is contained in the sets $\leftq{w}{L}$, for $w\in
L$. At the abstract level, we obviously have access to
$\leftq{h(w)}{h(L)}$, but we really need $h(\leftq{w}{L})$ in order
to ensure that properties satisfied within fairness by the abstraction
will also be satisfied within fairness by the concrete system in
a corresponding way.  Thus, we need to investigate the
relation between the sets $h(\leftq{w}{L})$ and $\leftq{h(w)}{h(L)}$
and find conditions under which $\leftq{h(w)}{h(L)}$ can be used
instead of $h(\leftq{w}{L})$.

In general, $h(\leftq{w}{L})$ is a proper subset of
$\leftq{h(w)}{h(L)}$. In order to obtain sufficient information about
$h(\leftq{w}{L})$ from $\leftq{h(w)}{h(L)}$, one would be tempted to
require equality of the two sets. Those homomorphisms are
\emph{continuation closed}, since computing the continuation or the
abstraction first, both have the same result. However, this is
stronger than needed. Indeed, since we are dealing with satisfaction
within fairness, we will show that it is sufficient that the behaviors in
$\leftq{h(w)}{h(L)}$ ``eventually'' become behaviors in
$h(\leftq{w}{L})$. This condition is the one called \em simplicity\em\
of an abstraction homomorphism in \cite{ochsenschlaeger94a}. We will
use a name that is more intuitive with respect to their definition and
call them \emph{weakly contiunation-closed} homomorphisms. Their exact
definition is the following.

\begin{definition}
An abstraction homomorphism $h:\Sigma^\infty\rightarrow\Sigma'^\infty$ is 
\emph{weakly continuation-closed} for a language $L\subseteq\Sigma^*$
and a word $w\in L$ if and only if there exists $u\in
\leftq{h(w)}{h(L)}$ such that $\leftq{u}{\leftq{h(w)}{h(L)}} =
\leftq{u}{h(\leftq{w}{L})}$. The homomorphism 
$h$ is \emph{weakly continuation-closed} for $L$ if and only if it is
for all words $w\in L$.
\end{definition}

Theorem~\ref{retrans} will show that this definition indeed meets all
the requirements we  have informally described above. More details
about weakly continuation-closed homomorphisms can be found in 
\cite{ochsenschlaeger94a}.

\section{Preservation of Linear Properties}

Before turning to the preservation of properties satisfied within fairness
by weakly continuation-closed homomorphisms, we need some general
results about abstraction homomorphisms and properties. The problem we
address is that the properties true of the abstracted system and of
the concrete system can rarely be identical. Indeed, one needs to take
into account the fact that the abstraction can rename or hide
symbols. Our goal here is to define a transformation on properties
that mirrors this.

We consider properties defined by PLTL formulas (see
Section~\ref{sec-prelim}). In order to make the definition of property
transformations easier and to make the interpretation of formulas over
words more direct (remember that we are dealing with systems
represented by sets of infinite words), we define some normal
forms for PLTL formulas.

A first restriction is to consider only positive normal form formulas.
\begin{definition}\label{pos-norm}
A PLTL-formula $\eta$ is in \em positive normal form \em if and only if
the scope of all negations is a single atomic proposition. 
\end{definition}

Now we turn to the problem of interpreting formulas over words.  Our
generic way of doing this (see Section~\ref{sec-prelim}) is to use a mapping
$\lambda: \Sigma \rightarrow 2^{AP}$ from the alphabet $\Sigma$ of the
word to the subsets of the atomic propositions $AP$ of the formula.
However, in this context, it is quite natural to consider the
elements of $\Sigma$ directly as atomic propositions, which implies
that one is using a mapping $\lambda_\Sigma$ such that $\forall a\in\Sigma:
\lambda_\Sigma(a)=\{a\}$. We define a normal form that corresponds to this.

\begin{definition}
Let $\Sigma$ be an alphabet.
We say that a PLTL formula $\eta$ is in \em $\Sigma$-normal form \em
if and only if $\eta$ is in positive normal form  and all its atomic
propositions are in $\Sigma$ (i.e. $AP\subseteq\Sigma$). 

For an alphabet $\Sigma$, the \em canonical $\Sigma$-labeling
function \em $\lambda_\Sigma: \Sigma\rightarrow 2^\Sigma$ is the one
such that $\forall a\in\Sigma:
\lambda_\Sigma(a)=\{a\}$.
\end{definition}

Note that using $\Sigma$-normal form formulas is not really
restrictive. Indeed, for any PLTL-formulas $\eta$ over a set $AP$ of 
atomic proposition and any labeling function $\lambda:\Sigma\rightarrow
2^{AP}$, there exists a PLTL-formula $\eta'$ in $\Sigma$-normal form such that,
for all $x\in\Sigma^\omega$, $x, \lambda\models\eta$ 
if and only if $x, \lambda_\Sigma \models\eta'$.

We now turn to the interaction between properties and abstraction
homomorphisms. Consider an abstraction homomorphism
$h:\Sigma^\infty\rightarrow\Sigma'^\infty$ and assume we have
established a ($\Sigma'$-normal form) property $\eta$ of the
abstract version  $L'_{\omega}\subseteq\Sigma'^\omega$ of a system 
obtained under this
homomorphism. Of what system can we say that the property is true on
the concrete level? One would expect $h^{-1}(L'_{\omega})$. However,
this is a language on $\Sigma$ on which we cannot directly interpret
$\eta$. One could modify $\eta$ to take this into account, but it is
simpler to modify the labeling function. 

\begin{definition}
For alphabets $\Sigma$ and $\Sigma'$, and for an abstraction
homomorphism $h:\Sigma^\infty\rightarrow\Sigma'^\infty$, the \em
canonical $h_{\Sigma\Sigma'}$-labeling function \em
$\lambda_{h_{\Sigma\Sigma'}} : \Sigma\rightarrow 2^{\Sigma' \cup
\{{\varepsilon}\}}$ is the one such that such that $\forall a\in\Sigma:
\lambda_{h_{\Sigma\Sigma'}}(a)=\{h(a)\}$.
\end{definition}

Notice that this labeling function maps some letters to the
proposition $\varepsilon$ which stands for the empty word. So, we
can't expect a formula $\eta$ true of the abstract system
$L'_{\omega}$ to be true of $h^{-1}(L'_{\omega})$, even using the
mapping $\lambda_{h_{\Sigma\Sigma'}}$. Indeed, this mapping takes care
of the fact that letters are renamed, but does not take care of the
fact that $\varepsilon$ is the empty word. What is needed is to ignore
the empty word in the evaluation of the formula.  This is handled by
transforming the formula $\eta$ from $\Sigma'$-normal form to
$\Sigma'\cup {\varepsilon}$-normal form as follows.


\begin{definition}\label{retrans-func-def}
Let $\eta$ be a PLTL-formula in $\Sigma'$-normal form. We define
recursively a mapping $T(\eta)$ that yields a formula in $\Sigma'\cup
{\varepsilon}$-normal form (see Figure~\ref{T-def}; $\hat b$ designates binary boolean 
operators: $\hat b\in\{\wedge,\vee,\Rightarrow,\Leftrightarrow\}$).
\begin{figure}[bth]
{\footnotesize
\[T(\eta)= \left\{
\begin{array}{ll}
true ,   & \mbox{if $\eta=true$,}
\bigskip \\
\neg(true) ,   & \mbox{if $\eta=\neg(true)$,}
\bigskip \\
a ,   & \mbox{if $\eta=a\in\Sigma'$,}
\bigskip \\
(\neg(a))\land (\neg(\varepsilon)) ,   
& \mbox{if $\eta=\neg(a)$ and $a\in\Sigma'$,}
\bigskip \\
(T(\xi))\,\hat{b}\,(T(\zeta)) , 
& \mbox{if $\eta=(\xi)\,\hat{b}\,(\zeta)$,} \bigskip \\
((\varepsilon)\vee (T(\xi)))\,\until\,(T(\zeta)) , 
& \mbox{if $\eta=(\xi)\until(\zeta)$,} \bigskip \\         
(T(\xi))\,\before\,(T(\zeta)) , 
& \mbox{if $\eta=(\xi)\before(\zeta)$,} \bigskip \\
\eventually(T(\xi)) , 
& \mbox{if $\eta=\eventually(\xi)$,} \bigskip \\
\always((\varepsilon)\vee(T(\xi))) ,
& \mbox{if $\eta=\always(\xi)$,} \bigskip \\
(\eps)\,\until((\neg(\eps))\,\wedge\,(\nxt((\eps)\,\until(T(\xi))))) , 
& \mbox{if $\eta=\nxt(\xi)$.}
\end{array}
\right. \]
}
\caption{The syntactical transformation of PLTL.}\label{T-def}
\end{figure}

As defined, the mapping $T$ does not modify pure Boolean formulas (not
including any temporal operator).
However, a pure Boolean formula $\eta$ should be mapped to
$(\varepsilon)\until(N(\eta))$ where $N$ replaces all
subformulas $\neg(a)$ of a PLTL-formula such that $a$ is an atomic proposition
by $(\neg(a))\wedge(\neg(\eps))$.
We thus extend $T$ into a mapping $R$ such that $R(\eta)$ is $T(\eta)$ with all maximal pure Boolean
subformulas  $\xi_b$ replaced by $(\varepsilon)\until(N(\xi_b))$.
\end{definition}

We can now give a statement relating a property true on an abstraction
of a system to a property true at the concrete level~\cite{nitsche94a,nitschediss}

\begin{lemma} \label{lemma-PLTL}
Let $L_\omega\subseteq\Sigma'^\omega$, let $\eta$ be a PLTL-formula 
in $\Sigma'$-normal form, and let $h:\Sigma^\infty\rightarrow\Sigma'^\infty$ 
be an abstraction homomorphism. Then
\[L'_\omega, \lambda_{\Sigma'}\models\eta\ \iff\ h^{-1}(L'_\omega),
\lambda_{h_{\Sigma\Sigma'}}\models R(\eta).\]
\end{lemma}

The proof of Lemma~\ref{lemma-PLTL} consist of two lemmas handling boolean 
formulas and purely temporal formulas respectively.

\begin{lemma}\label{lemma-boole}
Let $h:\Sigma^\infty\rightarrow\Sigma'^\infty$ be an abstraction  
homomorphism.  Let $x'\in\Sigma'^\omega$ be an abstract
computation and let $x\in h\inv(x')$. Let
$\eta$ be a boolean formula in $\Sigma'$-normal form. Then
\[x',\lambda_{\Sigma'}\models\eta\ \ \mbox{if and only if}\ \ x,\lambda_{h_{\Sigma\Sigma'}}\models(\varepsilon)\until(N(\eta)).\]
\end{lemma}

\begin{proof}
Let $i\in I\!\!N$ such that $h(x_i)=x'_1$ and, for all $j<i$, $h(x_j)=\varepsilon$. 
We have, for all atomic propositions $a\in\Sigma'$, that
$x',\lambda_{\Sigma'}\models a$ if and only if
$x_{(i...)},\lambda_{h_{\Sigma\Sigma'}}\models a$, and thus
$x',\lambda_{\Sigma'}\models\neg(a)$ if and only if
$x_{(i...)},\lambda_{h_{\Sigma\Sigma'}}\models\neg(a)$.
Because $h(x_i)\neq\eps$, we have $x',\lambda_{\Sigma'}\models\neg(a)$ if and only if
$x_{(i...)},\lambda_{h_{\Sigma\Sigma'}}\models(\neg(a))\wedge(\neg(\eps))$. According to the 
semantics of boolean connectives we obtain 
$x',\lambda_{\Sigma'}\models\eta$ if and only if
$x_{(i...)},\lambda_{h_{\Sigma\Sigma'}}\models N(\eta)$.

For all $j<i$, $h(x_j)=\varepsilon$, which means that 
$x\suf{j},\lambda_{h_{\Sigma\Sigma'}}\not\models N(\eta)$ and
$x_{(j...)},\lambda_{h_{\Sigma\Sigma'}}\models\varepsilon$. Thus $x',\lambda_{\Sigma'}\models\eta$
if and only if $x,\lambda_{h_{\Sigma\Sigma'}}\models(\varepsilon)\until(N(\eta))$. \qed
\end{proof}

\begin{lemma}\label{T-lemma}
Let $h:\Sigma^\infty\rightarrow\Sigma'^\infty$ be an abstraction 
homomorphism. Let $x'\in\Sigma'^\omega$ be an abstract
computation and let $x\in h\inv(x')$. Let
$\eta$ be a PLTL-formula in $\Sigma'$-normal form such that all atomic 
propositions are in the scope of a temporal operator (we call these formulas
\emph{purely temporal}). Then
\[x',\lambda_{\Sigma'}\models\eta\ \ \mbox{if and only if}\ \ x,\lambda_{h_{\Sigma\Sigma'}}\models T(\eta).\]
\end{lemma}

\begin{remark}
Lemma~\ref{T-lemma} is not surprising, because $T(\eta)$
takes care of subwords of $\omega$-words in
$h^{-1}(x')$ that $h$ takes to $\eps$, not changing the 
general structure of $\eta$. However, because many cases need to be 
distinguished, the proof of Lemma~\ref{T-lemma} is quite lengthy.
\end{remark}

\begin{proof}
The proof is by induction on the
structure of $\eta$. If $\eta$ contains exactly one 
temporal operator that quantifies over all
atomic propositions in $\eta$ (the induction's basis), then 
all proper subformulas $\xi$ of $\eta$ are 
boolean formulas and hence $T(\xi)=N(\xi)$.

By Lemma~\ref{lemma-boole} and since $T(\xi)=N(\xi)$, for all proper
subformulas $\xi$ of $\eta$ and all $x\in h^{-1}(x')$ we have 
$x',\lambda_{\Sigma'}\models\xi$ if and only if
$x,\lambda_{h_{\Sigma\Sigma'}}\models(\eps)\until(T(\xi))$. Therefore, if $h(x_1)\neq\eps$, 
$x',\lambda_{\Sigma'}\models\xi$ if and only if $x,\lambda_{h_{\Sigma\Sigma'}}\models T(\xi)$. We use this equivalence to
prove the induction's basis. Because all atomic propositions of $\eta$ are
in the scope of the only temporal operator, we need not prove the induction's 
basis for boolean connectives.

\underline{$\eta=(\xi)\until(\zeta)$:} 
$x',\lambda_{\Sigma'}\models(\xi)\until(\zeta)$ if and only if there exists
$i\in\nat$ such that $x'_{(i...)},\lambda_{\Sigma'}\models\zeta$ and
$x'_{(j...)},\lambda_{\Sigma'}\models\xi$, for all $j<i$. This is
equivalent to the existence of a
$k\in\nat$ such that $x\suf{k},\lambda_{h_{\Sigma\Sigma'}}\models
T(\zeta)$ and $x\suf{l},\lambda_{h_{\Sigma\Sigma'}}\models T(\xi)$,
for all $l<k$ such that $h(x_l)\neq\eps$. Thus, 
$x',\lambda_{\Sigma'}\models(\xi)\until(\zeta)$ if and only if 
$x,\lambda_{h_{\Sigma\Sigma'}}\models((\eps)\vee(T(\xi)))\until(T(\zeta))$.

\underline{$\eta=(\xi)\before(\zeta)$:} 
$x',\lambda_{\Sigma'}\models(\xi)\before(\zeta)$ if and only if there exists no $i\in\nat$ 
such that
$x'_{(i...)},\lambda_{\Sigma'}\models\zeta$ or there exists 
an $i\in\nat$ and a $j<i$ such that 
$x'_{(i...)},\lambda_{\Sigma'}\models\zeta$, 
$x'_{(j...)},\lambda_{\Sigma'}\models\xi$, and, for all $k<i$, 
$x'\suf{k},\lambda_{\Sigma'}\not\models
\zeta$. This is equivalent to: There exists no $l\in\nat$ such that 
$x\suf{l},\lambda_{h_{\Sigma\Sigma'}}\models T(\zeta)$, or there exists an $l\in\nat$ and an $m<l$ such that
$x\suf{l},\lambda_{h_{\Sigma\Sigma'}}\models T(\zeta)$, $x\suf{m},\lambda_{h_{\Sigma\Sigma'}}\models T(\xi)$, and, for all $n<l$, 
$x\suf{n},\lambda_{h}\not\models T(\xi)$. Therefore, $x',\lambda_{\Sigma'}\models(\xi)\before(\zeta)$ if 
and only if $x,\lambda_{h_{\Sigma\Sigma'}}\models(T(\xi))\before(T(\zeta))$.

\underline{$\eta=\eventually(\xi)$:}
$x',\lambda_{\Sigma'}\models\eventually(\xi)$ if and only if there exists 
$i\in\nat$ such that 
$x'_{(i...)},\lambda_{\Sigma'}\models\xi$. This is equivalent to the existence of 
$j\in\nat$ such that $x\suf{j},\lambda_{h_{\Sigma\Sigma'}}\models T(\xi)$. Hence, 
$x',\lambda_{\Sigma'}\models\eventually(\xi)$ if 
and only if $x,\lambda_{h_{\Sigma\Sigma'}}\models\eventually(T(\xi))$.

\underline{$\eta=\always(\xi)$:}
$x',\lambda_{\Sigma'}\models\always(\xi)$ if and only if $x'_{(i...)},\lambda_{\Sigma'}\models\xi$, 
for all $i\in\nat$. This is equivalent to: For all $j\in\nat$ with 
$h(x_j)\neq\eps$ we have $x\suf{j},\lambda_{h_{\Sigma\Sigma'}}\models
T(\xi)$. Since $h(x)=x'$
there are infinitely many different $j\in\nat$ with $h(x_j)\neq\eps$
and consequently 
$x',\lambda_{\Sigma'}\models\always(\xi)$ if 
and only if $x,\lambda_{h_{\Sigma\Sigma'}}\models\always((\eps)\vee(T(\xi)))$.

\noindent\underline{$\eta=\nxt(\xi)$:}
$x',\lambda_{\Sigma'}\models\nxt(\xi)$ if and only if $x'_{(2...)},\lambda_{\Sigma'}\models\xi$.
Equivalently, there exists a $j\in\nat$ and a $k<j$ such that $x\suf{j},\lambda_{h_{\Sigma\Sigma'}}\models
T(\xi)$, $h(x_k)\neq\eps$, and $h(x_l)=\eps$, for all $l<j$ such that
$l\neq k$. So $x',\lambda_{\Sigma'}\models\nxt(\xi)$ if and only if
$x,\lambda_{h_{\Sigma\Sigma'}}\models
(\eps)\until((\neg(\eps))\wedge(\nxt((\eps)\until(T(\xi)))))$.

This last step finishes the proof of the induction's basis. In the
inductive step, the proper subformulas of $\eta$ need not necessarily
satisfy the preconditions
of the lemma, because they can contain atomic 
propositions that are not in the scope of a temporal operator 
(of the subformula). Hence, in general, a subformula $\xi$ of $\eta$
is the boolean combination of boolean formulas $\xi_b$
and purely temporal formulas $\xi_t$. By induction, we
have
$x',\lambda_{\Sigma'}\models\xi_t$ if and only if $x,\lambda_{h_{\Sigma\Sigma'}}\models T(\xi_t)$.
According to Lemma~\ref{lemma-boole}, $x',\lambda_{\Sigma'}\models\xi_b$ if and only if 
$x,\lambda_{h_{\Sigma\Sigma'}}\models (\eps)\until(N(\xi_b))$. Thus $x',\lambda_{\Sigma'}\models\xi_b$ if and only if 
$x,\lambda_{h_{\Sigma\Sigma'}}\models (\eps)\until(T(\xi_b))$, because
$T(\xi_b)=N(\xi_b)$. Hence, if $h(x_1)\neq\eps$, then
$x',\lambda_{\Sigma'}\models\xi_b$ 
if and only if $x,\lambda_{h_{\Sigma\Sigma'}}\models T(\xi_b)$. Therefore, for all subformulas $\xi$ of $\eta$,
we have: if 
$h(x_1)\neq\eps$, then $x',\lambda_{\Sigma'}\models\xi$ if and only if $x,\lambda_{h_{\Sigma\Sigma'}}\models
T(\xi)$. We use this condition as our induction's hypothesis.

\underline{$\eta=(\xi)\hat b(\zeta)$:} 
Because of the lemma's preconditions, $\xi$ and $\zeta$ must be purely
temporal subformulas of $\eta$, for a binary boolean connective $\hat
b$. Then, by induction and the semantics of boolean connectives,
$x',\lambda_{\Sigma'}\models(\xi)\hat b(\zeta)$ 
if and only if $x,\lambda_{h_{\Sigma\Sigma'}}\models(T(\xi))\hat b(T(\zeta))$.

\underline{$\eta=(\xi)\until(\zeta)$:} 
$x',\lambda_{\Sigma'}\models(\xi)\until(\zeta)$ if and only if there exists $i\in\nat$ 
such that $x'_{(i...)},\lambda_{\Sigma'}\models\zeta$ and, for all $j<i$, 
$x'_{(j...)},\lambda_{\Sigma'}\models\xi$. By induction, 
this is equivalent to the existence
of $k\in\nat$ such that $x\suf{k},\lambda_{h_{\Sigma\Sigma'}}\models
T(\zeta)$, and, for all $l<k$ we have
$x\suf{l},\lambda_{h_{\Sigma\Sigma'}}\models T(\xi)$ or $h(x_l)=\eps$.
Therefore, $x',\lambda_{\Sigma'}\models(\xi)\until(\zeta)$ if and only if 
$x,\lambda_{h_{\Sigma\Sigma'}}\models((\eps)\vee(T(\xi)))\until(T(\zeta))$.

\underline{$\eta=(\xi)\before(\zeta)$:} 
$x',\lambda_{\Sigma'}\models(\xi)\before(\zeta)$ if and only if there exists no $i\in\nat$ 
such that
$x'_{(i...)},\lambda_{\Sigma'}\models\zeta$ or there exists 
an $i\in\nat$ and a $j<i$ such that 
$x'_{(i...)},\lambda_{\Sigma'}\models\zeta$, 
$x'_{(j...)},\lambda_{\Sigma'}\models\xi$, and
$x'\suf{k},\lambda_{\Sigma'}\not\models \zeta$, for all $k<i$. By
induction, this 
is equivalent to: There exists no $l\in\nat$ such that 
$x\suf{l},\lambda_{h_{\Sigma\Sigma'}}\models T(\zeta)$ or there exists an $l\in\nat$ and an $m<l$ such that
$x\suf{l},\lambda_{h_{\Sigma\Sigma'}}\models T(\zeta)$,
$x\suf{m},\lambda_{h_{\Sigma\Sigma'}}\models T(\xi)$, and
$x\suf{n},\lambda_{h}\not\models T(\xi)$, for all $n<l$. Therefore,
$x',\lambda_{\Sigma'}\models(\xi)\before(\zeta)$ if 
and only if $x,\lambda_{h_{\Sigma\Sigma'}}\models(T(\xi))\before(T(\zeta))$.

\underline{$\eta=\eventually(\xi)$:}
$x',\lambda_{\Sigma'}\models\eventually(\xi)$ if and only if there exists 
$i\in\nat$ such that 
$x'_{(i...)},\lambda_{\Sigma'}\models\xi$. By induction, this 
is equivalent to the existence of 
$j\in\nat$ such that $x\suf{j},\lambda_{h_{\Sigma\Sigma'}}\models T(\xi)$. Hence, 
$x',\lambda_{\Sigma'}\models\eventually(\xi)$ if 
and only if $x,\lambda_{h_{\Sigma\Sigma'}}\models\eventually(T(\xi))$.

\underline{$\eta=\always(\xi)$:}
$x',\lambda_{\Sigma'}\models\always(\xi)$ if and only if $x'_{(i...)},\lambda_{\Sigma'}\models\xi$, 
for all $i\in\nat$. By induction, this is equivalent to: 
For all $j\in\nat$ such that 
$h(x_j)\neq\eps$, we have $x\suf{j},\lambda_{h_{\Sigma\Sigma'}}\models T(\xi)$. Since
$h(x)=x'$, there are infinitely  many
different $j\in\nat$ such that $h(x_j)\neq\eps$. Therefore 
$x',\lambda_{\Sigma'}\models\always(\xi)$ if 
and only if $x,\lambda_{h_{\Sigma\Sigma'}}\models\always((\eps)\vee(T(\xi)))$.

\underline{$\eta=\nxt(\xi)$:}
$x',\lambda_{\Sigma'}\models\nxt(\xi)$ if and only if $x'_{(2...)},\lambda_{\Sigma'}\models\xi$.
Equivalently, by induction, there exists 
$j\in\nat$ and $k<j$ such that $x\suf{j},\lambda_{h_{\Sigma\Sigma'}}\models T(\xi)$, 
$h(x_k)\neq\eps$, and $h(x_l)=\eps$, for all $l<j$ such that $l\neq k$.
So, $x',\lambda_{\Sigma'}\models\nxt(\xi)$ if and only if $x,\lambda_{h_{\Sigma\Sigma'}}\models
(\eps)\until((\neg(\eps))\wedge(\nxt((\eps)\until(T(\xi)))))$. \qed
\end{proof}


\begin{proof}[of Lemma~\ref{lemma-PLTL}]
Lemma~\ref{lemma-boole}
and Lemma~\ref{T-lemma} establish the result. \qed
\end{proof}

\section{Preservation of Properties Satisfied within Fairness}
\label{sec-abstract}

Let $L\subseteq\Sigma^*$ be a prefix-closed language, let
$h:\Sigma^\infty\rightarrow\Sigma'^\infty$ be an abstraction
homomorphism, and let $\eta$ be a PLTL-formula in
$\Sigma'$-normal form. Assume that $\eta$ is satisfied by $lim(h(L))$
within fairness; in our notation $lim(h(L)), \lambda_{\Sigma'} \modls
\eta$. We will prove that, if the homomorphism $h$ is weakly
continuation-closed, then the property corresponding to $\eta$ is also
satisfied within fairness by $lim(L)$, i.e.\ that $lim(L),
\lambda_{h_{\Sigma\Sigma'}} \modls R(\eta)$.
To establish this result we need a condition that allows to commute
Eilenberg-limit and homomorphism application.

\begin{lemma}\label{corollary5}
If $L\subseteq\Sigma^{*}$ is a prefix-closed regular language and 
$h: \Sigma^\infty\rightarrow\Sigma'^\infty$ is an abstraction homomorphism, 
then $lim(h(L))=h(lim(L))$.
\end{lemma}

Lemma~\ref{corollary5} appears to be rather trivial. But, in fact, it
neither holds for regular languages that are not prefix-closed nor for 
prefix-closed languages that are not regular. The languages $a^*\cdot b$
and $pre(\{b^i\cdot a^i\mid i \in \nat\})$ reveal this observation for the 
homomorphism defined by $h(a)=a$ and $h(b)=\eps$. 
To prove the lemma, we use K\"onig's Lemma in a suitable version 
(\cite{hoogeboomrozenberg86}, Lemma 3.3.):

\begin{lemma}[K\"onig's Lemma]
Let ${\cal R}\subseteq E\times E$ be a relation --- $E$ is an arbitrary 
set --- and
let, for all $n\in \nat$, $E_n$ be a finite nonempty subset of $E$ such that
$\bigcup_{n\in \nat} E_n$ is infinite and to
each $e\in E_{n+1}$ there exists an $f\in E_n$ such that $(f,e)\in\cal R$.
Then there exists an infinite sequence $(e_n)_{n\in \nat}$ in $E$ such that
$e_n\in E_n$ and $(e_n,e_{n+1})\in\cal R$ for all $n\in \nat$.
\end{lemma}

\begin{proof}[of Lemma~\ref{corollary5}]
``$lim(h(L))\subseteq h(lim(L))$'': 
We assume $lim(h(L))\neq\emptyset$ (otherwise the condition holds trivially).

If $x$ is an 
$\omega$-word in $lim(h(L))$, then $pre(x)\subseteq h(L)$ (remember
that $L$ and therefore $h(L)$ are prefix-closed). Let $w^n$ be the
prefix of $x$ of length $n$.\footnote{The notation $w^n$ should not be confused with
the $n$th power of $w$ ($n$ is just an index).} $(w^n)_{n\in\nat}$ is
then the
sequence of all prefixes of $x$ and thus generates $x$ as its limit.

To each of the $w^n$ we construct a set $U_n$ of \emph{minimal} inverse images of 
$w^n$. Let $U_n$ be the set of all words $u$ in $h^{-1}(w^n)\cap L$, such
that there is no shorter word $v$ in $h^{-1}(w^n)\cap L$ with $\leftq{u}{L}=
\leftq{v}{L}$. We define
\[U_n=\{u\in h^{-1}(w^n)\cap L\mid \not\exists v\in h^{-1}(w^n)\cap L: 
|u|>|v|\,\wedge\,\leftq{u}{L}=\leftq{v}{L}\}.\]

Because all $w^n$ are in $h(L)$ there must be a
$u\in L$ such that $h(u)=w^n$ to each $w^n$. Consequently, 
$U_n$ is not empty, for all $n\in \nat$.

Let $u\in U_n$. For all $v\in U_n$ such that
$\leftq{u}{L}=\leftq{v}{L}$, we have $|v|=|u|$ by definition of $U_n$. Because the set $\{\leftq{t}{L}\mid t\in\Sigma^*\}$ is finite (its cardinality
corresponds to the number of states in the minimal automaton accepting
$L$), we obtain:
$U_n$ is a finite set, for all $n\in \nat$.

Because $U_n\cap U_m=\emptyset$ if $n\neq m$ and all $U_n$
are nonempty sets, we observe that
$\bigcup_{n\in \nat} U_n$ is an infinite set.

By $\prec$ we denote the proper prefix relation; i.e.\ for all $u, v\in
\Sigma^*$, $u\prec v$ if and only if $u\neq v$ and $u\in pre(v)$. We show:
For all $n\in \nat$ and all $v\in U_{n+1}$, there exists a word $u\in U_n$
such that $u\prec v$.
Let $v$ be in $U_{n+1}$ and let $u$ be in $pre(v)$ such that $h(u)=w^n$. 
Hence $u\prec v$. Because $L$ is prefix-closed, $u$ is in $L$ and thus 
$u\in h^{-1}(w^n)\cap L$. The remainder of $v$ after $u$ we call $v'$; 
i.e.\ $v=uv'$. We assume that $u$ is not in $U_n$ and show a contradiction.

If $u\not\in U_n$, then there must be a word
$u'\in h^{-1}(w^n)\cap L$ such that $|u'|<|u|$ and
$\leftq{u}{L}=\leftq{u'}{L}$. Because $u'$ is in $h^{-1}(w^n)\cap L$,
we have $h(u'v')=w^{n+1}$. Because $\leftq{u}{L}=\leftq{u'}{L}$, we
obtain $u'v'\in L$ and $\leftq{v}{L}=\leftq{(u'v')}{L}$.
So $u'v'$ is in $h^{-1}(w^{n+1})\cap L$, $\leftq{(u'v')}{L}=\leftq{v}{L}$ and
$|u'v'|<|v|$. Therefore $v\not\in U_{n+1}$, which contradicts the choice of
$v$.

Hence all preconditions to apply K\"onig's Lemma are satisfied by the sets
$U_n$, $n\in\nat$, and thus there exists an infinite sequence
$(u^n)_{n\in \nat}$ of words in $L$ such that $u^n\in U_n$ and
$u^n\prec u^{n+1}$, for all $n\in \nat$. The sequence $(u^n)_{n\in \nat}$ 
uniquely generates some $y\in lim(L)$ and, because $h(u^n)=w^n$, for
all $n\in\nat$, we obtain $h(y)=x$. So, for all $x\in lim(h(L))$,
there exists a $y\in lim(L)$ such that $x=h(y)$. Thus $lim(h(L))\subseteq h(lim(L))$. 

``$h(lim(L))\subseteq lim(h(L))$'': 
Let $h(lim(L))\neq\emptyset$.
Let $x$ be in $lim(L)$, such that 
$h(x)$ is defined. Because $L$
is prefix-closed, all $u\in pre(x)$ are in $L$.
So, for all $u\in pre(x)$, $h(u)$ is 
in $pre(h(x))$. Because $h(x)$ is defined, there are 
infinitely many different $h(u)$ in $pre(h(x))$, for $u\in pre(x)\subseteq L$. 
Thus $h(x)$ is in $lim(h(L))$, and we obtain $h(lim(L))\subseteq
lim(h(L))$. \qed
\end{proof}

Using Lemma~\ref{corollary5}, we can now prove a result relating 
a property satisfied within fairness by $lim(h(L))$ to a property
satisfied within fairness by $lim(L)$.

\begin{theorem}
\label{retrans-cor}
Let $L\subseteq\Sigma^{*}$ be a prefix-closed regular language,
let $h: \Sigma^\infty\rightarrow\Sigma'^\infty$ be an abstraction homomorphism 
such that $h$ is weakly continuation-closed on $L$ and $h(L)$ does not contain 
maximal words\footnote{Maximal words in $h(L)$ are words that
are not a proper prefix of another word in $h(L)$. We will lift the 
restriction to maximal-word-free abstractions in the next section.}, and
let $\eta$ be a PLTL-formula in $\Sigma'$-normalform. Then
\[lim(h(L)),\lambda_{\Sigma'}\modls\eta\ \ \ \mbox{if and only if}\ \ \ 
lim(L),\lambda_{h_{\Sigma\Sigma'}}\modls R(\eta).\]
\end{theorem}

This theorem is will be a consequence of the following two lemmas
(Lemma~\ref{retrans} and Lemma~\ref{retrans2}).

\begin{lemma}
\label{retrans}
Let $L\subseteq\Sigma^{*}$ be a prefix-closed regular language,
let $h: \Sigma^\infty\rightarrow\Sigma'^\infty$ be an abstraction homomorphism 
such that $h$ is weakly continuation-closed on $L$ and $h(L)$ does not
contain maximal words, and
let $\eta$ be a PLTL-formula in $\Sigma'$-normal form. We have that
\[lim(h(L)), \lambda_{\Sigma'}\modls \eta\ \ \ \mbox{implies}\ \ \ 
lim(L), \lambda_{h_{\Sigma\Sigma'}} \modls R(\eta).\]
\end{lemma}

\setcounter{equation}{0}
\begin{proof}
We assume that $lim(h(L)), \lambda_{\Sigma'}\modls \eta$ and 
derive  $lim(L), \lambda_{h_{\Sigma\Sigma'}} \modls
R(\eta)$. 
By definition $lim(L), \lambda_{h_{\Sigma\Sigma'}} \modls R(\eta)$ if for all $u\in L$, there exists some $x\in\leftq{u}{lim(L)}$ such that 
$ux,\lambda_{h_{\Sigma\Sigma'}} \models R(\eta)$. Consider thus an
arbitrary $u \in L$. Because $h$ is weakly continuation-closed on $L$, there exists  
$v\in \leftq{h(u)}{h(L)}$ such that 
\[\leftq{v}{h(\leftq{u}{L})}=\]
\begin{equation}\label{condition1}
\leftq{v}{\leftq{h(u)}{h(L)}}=
\end{equation}
\[\leftq{h(u)v}{h(L)}.\]

As $lim(h(L)), \lambda_{\Sigma'}\modls \eta$, we
get $\forall r\in pre(lim(h(L))): \exists s\in
\leftq{r}{lim(h(L))}: rs ,\lambda_{\Sigma'}\models \eta$, and in
particular, by substituting $h(u)v$ for $r$, there exists some $y\in
\leftq{h(u)v}{lim(h(L))}=lim(\leftq{h(u)v}{h(L)})$ such that
\begin{equation}\label{condition2}
h(u)vy, \lambda_{\Sigma'}\models \eta.
\end{equation}
Given equation~(\ref{condition1}) this is equivalent to
\[y\in lim(\leftq{v}{h(\leftq{u}{L})})=\]
\[\leftq{v}{lim(h(\leftq{u}{L}))}.\]

Thus we know that $vy$ is in $lim(h(\leftq{u}{L}))$, which, in view of  
Lemma~\ref{corollary5}, is
equivalent to 
\[vy\in h(lim(\leftq{u}{L})).\] 

So, there exists 
$x\in lim(\leftq{u}{L})$ such that
\begin{equation}\label{condition3}
h(x)=vy.
\end{equation}

Viewing $vy$ as a single word $z$, we have shown that
for all $u\in L$, there exists $x\in lim(\leftq{u}{L})$ and $z\in
\leftq{h(u)}{lim(h(L))}$ such 
that $h(x)=z$ (because of equation (\ref{condition3})) and
$h(u)z, \lambda_{\Sigma'}\models \eta$ (because of equation (\ref{condition2})). 

Consider now the language $\tilde L=pre(ux)$ of prefixes of $ux$.
Clearly, $lim(\tilde L)=\{ux\}$ and $lim(h(\tilde L))=\{h(u)z\}$.

Because $h(u)z,\lambda_{\Sigma'}\models \eta$, we have
$lim(h(\tilde L)),\lambda_{\Sigma'}\models\eta$. 
Using Lemma~\ref{lemma-PLTL} and given that
$lim(\tilde L)\subseteq h^{-1}(lim(h(\tilde L)))$, we obtain 
$lim(\tilde L),\lambda_{h_{\Sigma\Sigma'}} \models R(\eta)$, 
or $ux,\lambda_{h_{\Sigma\Sigma'}} \models R(\eta)$.
We have thus shown that for all $u\in 
L$, there exists $x\in\leftq{u}{lim(L)}$, such 
that $ux,\lambda_{h_{\Sigma\Sigma'}} \models R(\eta)$.
Hence
we have shown that $lim(L), \lambda_{h_{\Sigma\Sigma'}} \modls R(\eta)$
\qed
\end{proof}

As discussed in Section~\ref{sec-example} using an example,
Lemma~\ref{retrans} does not hold, if we do not require the
abstraction homomorphism to be weakly continuation-closed.

\begin{lemma}
\label{retrans2}
Let $L\subseteq\Sigma^{*}$ be a prefix-closed regular language.
Let $h: \Sigma^\infty\rightarrow\Sigma'^\infty$ be an abstraction
homomorphism such that $h(L)$ does not contain maximal words. Let
$\eta$ be a PLTL-formula in $\Sigma'$-normalform.  Then
\[lim(L),\lambda_{h_{\Sigma\Sigma'}}\modls R(\eta)\ \ \ \mbox{implies}\ \ \
lim(h(L)),\lambda_{\Sigma'}\modls\eta.\]
\end{lemma}

\begin{proof}
We assume that $lim(L),\lambda_{h_{\Sigma\Sigma'}}\modls R(\eta)$ and show
that $lim(h(L)),\lambda_{\Sigma'}\modls\eta$.
Let $w'\in pre(lim(h(L)))$, let $w\in pre(lim(L))\cap h\inv(w')$, and
let $x\in\leftq{w}{lim(L)}$ such that 
$wx,\lambda_{h_{\Sigma\Sigma'}}\models R(\eta)$.

If $h(wx)$ is defined, then, by Lemma~\ref{lemma-PLTL},
$h(wx),\lambda_{\Sigma'}\models\eta$. Therefore, there exists an
$x'=h(x)\in\leftq{w'}{lim(h(L))}$ such that
$w'x',\lambda_{\Sigma'}\models\eta$.

If $h(wx)$ is undefined, then there is a prefix $v$ of $wx$ such
that $h(\leftq{v}{pre(wx)})=\{\eps\}$. (In fact, there are infinitely
many of these prefixes $v$.) Then, by definition of $R$ and
$\lambda_{h_{\Sigma\Sigma'}}$, we have, for all $y\in\Sigma^\omega$, that
$vy,\lambda_{h_{\Sigma\Sigma'}}\models R(\eta)$. 

If there exists 
$y\in\Sigma^\omega$ such that $h(y)\in\leftq{h(v)}{lim(h(L))}$, then
let $x'$ be the only $\omega$-word in
$\leftq{w'}{\{h(vy)\}}$. $x'$ is in $\leftq{w'}{lim(h(L))}$. 
So by Lemma~\ref{lemma-PLTL},
$w'x',\lambda_{\Sigma'}\models\eta$.

If there exists no $y\in\Sigma^\omega$ such that
$h(y)\in\leftq{h(v)}{lim(h(L))}$, then $h(L)$ contains maximal words,
which contradicts the theorem's preconditions.

So, for all $w'\in pre(lim(h(L)))$, there exists an
$x'\in\leftq{w'}{lim(h(L))}$ such that
$w'x',\lambda_{\Sigma'}\models\eta$. Thus
$lim(h(L)),\lambda_{\Sigma'}\modls\eta$. \qed
\end{proof}

We discuss in the next section how we can extend Theorem~\ref{retrans-cor} 
to deal with maximal words.
\section{Improving the Results}

If a language $L\subseteq\Sigma^*$ contains maximal words, i.e.\ words
that have no continuation in $L$, then $lim(L)$ contains no
information about them: if $w$ is a maximal word in
$L$, then $w\not\in pre(lim(L))$. To avoid this loss of information
we extend maximal words by dummy-letters. Formally, we define 
satisfaction within fairness on $L$ itself instead of
$lim(L)$.

\begin{definition}
Let $L\subseteq\Sigma^*$. Let $\#\not\in\Sigma$.
We define the set of maximal words of $L$ by $max(L)=\{w\in L\mid
\leftq{w}{L}=\{\eps\}\}$.
We define the extension of $L$ to be $xtd(L)=L\cup
max(L)\cdot\{\#\}^*$.
\end{definition}

If $L$ is a regular language, then the construction of an automaton
accepting $xtd(L)$ is easy: for all accepting states in a reduced 
deterministic automaton for
$L$ that have no outgoing transition, add a self-loop labelled with
$\#$ to that state. Then the resulting automaton accepts $xtd(L)$.

\begin{definition}
Let $L\subseteq\Sigma^*$, let $\eta$ be a PLTL-formula, and let
$\lambda:\Sigma\rightarrow 2^{AP}$ be a labelling function. $L$
satisfies $\eta$ within fairness with respect to $\lambda$ (written:
``$L,\lambda\modls\eta$'') if and only if
$lim(xtd(L)),\lambda\modls\eta$.
\end{definition}

\begin{definition}\label{eps-ext}
Let $\Sigma$ be an alphabet.
A PLTL-formula is in extended $\Sigma$-normal form if and only if it is
in positive normal form (Definition~\ref{pos-norm}), $\Sigma\cup\{\eps\}$ is its set of atomic
propositions, and it contains the atomic proposition $\eps$ only in the
form $\always(\eps)$ (``all actions are hidden by the abstraction'').
\end{definition}

\begin{definition}
Let $\lambda:\Sigma\rightarrow 2^{AP}$ be a labelling function for an
alphabet $\Sigma$ and a set of atomic propositions $AP$. 

We
define the $\eps$-extension of $\lambda$ to be the function 
$\lambda^{\eps}:\Sigma\cup\{\#\}\rightarrow 2^{AP\cup\{\eps\}}$ such that 
$\lambda^{\eps}(a)=\lambda(a)$, for all $a\in\Sigma$, and
$\lambda^{\eps}(\#)=\{\eps\}$.

We
define the $\#$-extension of $\lambda$ to be the function 
$\lambda^\#:\Sigma\cup\{\#\}\rightarrow 2^{AP\cup\{\#\}}$ such that 
$\lambda^\#(a)=\lambda(a)$, for all $a\in\Sigma$, and $\lambda^\#(\#)=\{\#\}$.
\end{definition}

\begin{theorem}\label{retrans-numsign}
Let $h:\Sigma^\infty\rightarrow\Sigma'^\infty$ be a weakly continuation-closed
homomorphism on the prefix-closed
regular language 
$L\subseteq\Sigma^*$. Let $\eta$ be a PLTL-formula in extended
$\Sigma'$-normalform. Then
\[L,\lambda_h^{\eps}\modls R(\eta)\ \ \mbox{if and only if}\ \
h(L),\lambda_{\Sigma'}^{\eps}\modls\eta.\]
\end{theorem}

\begin{proof}
Let the extension of $L$ with respect to empty abstract
suffixes be the language $xtd_h(L)=L\cup\{w\in L\mid h(\leftq{w}{L})=\{\eps\}\}\cdot\{\#\}^*$.

Let
$h':(\Sigma\cup\{\#\})^\infty\rightarrow(\Sigma'\cup\{\#\})^\infty$ be
the abstraction homomorphism defined by $h'(a)=h(a)$, for all
$a\in\Sigma$, and $h'(\#)=\#$. Because $h$ is weakly
continuation-closed on $L$, $h'$ is weakly continuation-closed on
$xtd_h(L)$ and $h'(xtd_h(L))=xtd(h(L))$. The latter equality
holds, because $h$ being weakly continuation-closed on $L$ implies for
all $w\in L$, $\leftq{h(w)}{h(L)}=\{\eps\}$ if
$h(\leftq{w}{L})=\{\eps\}$ \cite{ochsenschlaeger92}. Because $h'(xtd_h(L))=xtd(h(L))$,
$h'(xtd_h(L))$ does not contain maximal words.

Let $\eta'$ be the PLTL-formula that we obtain by replacing the atomic
proposition $\eps$ in $\eta$ by a new atomic proposition $\#$. We have
\begin{itemize}
\item $lim(h'(xtd_h(L))),\lambda_{\Sigma'}^{\eps}\modls\eta$ if and only
if
$lim(h'(xtd_h(L))),\lambda_{\Sigma'}^{\#}\modls\eta'$,
\item $lim(xtd(L)),\lambda_h^{\eps}\modls R(\eta)$ if and only if
$lim(xtd_h(L)),\lambda_h^{\#}\modls R(\eta')$, and
\item $lim(xtd_h(L)),\lambda_h^{\#}\modls R(\eta')$ if and only
if $lim(xtd_h(L)),\lambda_{h'}\modls R(\eta')$.
\end{itemize}

Additionally, by Theorem~\ref{retrans-cor}, we have that $lim(h'(xtd_h(L))),\lambda_{\Sigma'}^{\#}\modls\eta'$ if and
only if $lim(xtd_h(L)),\lambda_{h'}\modls
R(\eta')$. According to the
above established equivalences and $h'(xtd_h(L))=xtd(h(L))$, we finally obtain 
$L,\lambda_h^{\eps}\modls R(\eta)$ if and only if
$h(L),\lambda_{\Sigma'}^{\eps}\modls\eta$. \qed
\end{proof}

If the above result is not restricted to PLTL properties but extended
to all possible $\omega$-languages as properties, one can also show that weak
continuation-closure of a homomorphism is not
only a sufficient but also a necessary condition for an abstraction to
preserve properties satisfied within fairness
\cite{nitschediss,nitscheochsenschlaeger96a}. 

\section{Conclusion}

We have introduced \emph{satisfaction within fairness} as a satisfaction
relation with an inherent abstract notion of fairness. It is defined
in terms of \emph{relative liveness properties}
\cite{alurhenzinger95,henzinger92}, lifted from a property
classification to a satisfaction relation
\cite{nitscheochsenschlaeger96a,nitschewolper97}. Besides exploring
the basic properties of the relation --- including exploring its dual,
\emph{relative safety} --- we have motivated its definition by considering
a small but typical introductory example of a distributed system.

We have established the link from satisfaction within fairness to the usual
satisfaction of linear-time properties under fairness by showing that,
to a regular system behavior satisfying a linear-time property within
fairness, a finite-state implementation can always be found that
satisfies the property under strong fairness. As the this 
finite-state implementation is usually significantly bigger (many more
states) than the most compact finite-state implementation of the
behavior, satisfaction within fairness offers a way of dealing with
linear-time satisfaction under fairness uisng more compact behavior
representations.

Since, however, state-spaces of realistic systems are far too large to
effectively be constructed, we have looked at \emph{behavior abstractions}
to decrease the size of the state space. Behavior abstraction is,
compared to abstract interpretation, a relatively primitive but by
that easy-to-apply approach to tackle state-space explosion. The two
concepts in behavior abstractions are action
\emph{renaming} and \emph{hiding}. These concepts can be defined in
terms of language homomorphisms extended to operate on
$\omega$-languages. In particular action renaming alters patterns of
events in computations of a system. To handle these
alterations on the level of linear-time temporal logic model-checking,
we use a syntactic transformation of PLTL-formulas. We show that an
abstract computation of the system satisfies a PLTL-formula if and
only if the concrete computation that results in the abstract one
satisfies the syntacticly transformed formula.

As discussed in the context of the motivating example
mentioned above, it turns out that behaviors abstractions are in
general too imprecise to preserve properties satisfied within
fairness. Here, preservation refers to a property being true on the
abstract level implying a corresponding property (the syntacticly
transformed one) being true on the concrete level. Elaborating on this we
give a condition for abstraction homomorphisms that guarantees the
preservation of properties satisfied within fairness by the
abstraction. The condition that abstraction homomorphisms must satisfy
is \emph{weak continuation-closure} \cite{ochsenschlaeger92}. The 
initial preservation result we
establish for weakly continuation-closed abstractions and properties
satisfied within fairness only holds for behaviors in which no
computation is finite (no maximal words in the language representing
the behavior). We have extended the result to capture also behaviors that
contain terminating computations.

For practical purposes \cite{nitsche98}, it is essential to be able to obtain a
representation of the abstract behavior of a system without an
exhaustive construction of the concrete one. It appears promising to
tackle this problem by applying partial-order reduction. The aim is to
construct a (partial-order) reduced state-space that results 
in the same abstract state-space as the concrete
state-space would. In addition, it must be possible to check weak
continuation-closure of the abstraction on the concrete state-space
by only considering the partial-order reduced one. A first major
result in that direction is presented in \cite{ultes-nitschestjames00},
where the persistent-set selective search
\cite{godefroidwolper93,wolpergodefroid93} partial-order technique is
applied in the context of the abstractions presented in this paper. The
efficient construction of abstract state spaces beyond 
\cite{ultes-nitschestjames00} as well as efficiently
checking weak continuation-closure will be topics for further study.

\bibliography{publ,refs}

\begin{thebibliography}{}

\bibitem[\protect\citeauthoryear{Abadi and Lamport}{Abadi and
  Lamport}{1988}]{abadilamport88}
{\sc Abadi, M.} {\sc and} {\sc Lamport, L.} 1988.
\newblock The existence of refinement mappings.
\newblock SRC Report~29, DEC System Research Center. July.

\bibitem[\protect\citeauthoryear{Abadi and Lamport}{Abadi and
  Lamport}{1990}]{abadilamport90}
{\sc Abadi, M.} {\sc and} {\sc Lamport, L.} 1990.
\newblock Composing specifications.
\newblock SRC Report~66, DEC System Research Center. October.

\bibitem[\protect\citeauthoryear{Alpern and Schneider}{Alpern and
  Schneider}{1985}]{alpernschneider85}
{\sc Alpern, B.} {\sc and} {\sc Schneider, F.~B.} 1985.
\newblock Defining liveness.
\newblock {\em Information Processing Letters\/}~{\em 21,\/}~4 (October),
  181--185.

\bibitem[\protect\citeauthoryear{Alur and Henzinger}{Alur and
  Henzinger}{1995}]{alurhenzinger95}
{\sc Alur, R.} {\sc and} {\sc Henzinger, T.~A.} 1995.
\newblock Local liveness for compositional modeling of fair reactive systems.
\newblock In {\em Computer Aided Verification (CAV) '95}, {P.~Wolper}, Ed.
  Lecture Notes in Computer Science, vol. 939. Springer, 166--179.

\bibitem[\protect\citeauthoryear{Berstel}{Berstel}{1979}]{berstel79}
{\sc Berstel, J.} 1979.
\newblock {\em Transductions and Context-Free Languages\/}, first ed.
\newblock Studienb\"ucher Informatik. Teubner Verlag, Stuttgart.

\bibitem[\protect\citeauthoryear{Eilenberg}{Eilenberg}{1974}]{eilenberg74}
{\sc Eilenberg, S.} 1974.
\newblock {\em Automata, Languages and Machines}. Vol.~A.
\newblock Academic Press, New York.

\bibitem[\protect\citeauthoryear{Emerson}{Emerson}{1990}]{emerson90}
{\sc Emerson, E.~A.} 1990.
\newblock Temporal and modal logic.
\newblock See \citeN{HTCS90B}, 995--1072.

\bibitem[\protect\citeauthoryear{Francez}{Francez}{1986}]{francez86}
{\sc Francez, N.} 1986.
\newblock {\em Fairness\/}, first ed.
\newblock Springer Verlag, New York.

\bibitem[\protect\citeauthoryear{Garey and Johnson}{Garey and
  Johnson}{1979}]{GJ79}
{\sc Garey, M.~R.} {\sc and} {\sc Johnson, D.~S.} 1979.
\newblock {\em Computers and Intractability. A Guide to the Theory of
  NP-Completeness}.
\newblock W.H. Freeman and Co., New York.

\bibitem[\protect\citeauthoryear{Godefroid and Wolper}{Godefroid and
  Wolper}{1993}]{godefroidwolper93}
{\sc Godefroid, P.} {\sc and} {\sc Wolper, P.} 1993.
\newblock Using partial orders for the efficient verification of deadlock
  freedom and safety properties.
\newblock {\em Formal Methods in System Design\/}~{\em 2,\/}~2 (April),
  149--164.

\bibitem[\protect\citeauthoryear{Harrison}{Harrison}{1978}]{harrison78}
{\sc Harrison, M.~A.} 1978.
\newblock {\em Introduction to Formal Language Theory\/}, first ed.
\newblock Addison-Wesley, Reading, Mass.

\bibitem[\protect\citeauthoryear{Henzinger}{Henzinger}{1992}]{henzinger92}
{\sc Henzinger, T.~A.} 1992.
\newblock Sooner is safer than later.
\newblock {\em Information Processing Letters\/}~{\em 43}, 135--141.

\bibitem[\protect\citeauthoryear{Hoogeboom and Rozenberg}{Hoogeboom and
  Rozenberg}{1986}]{hoogeboomrozenberg86}
{\sc Hoogeboom, H.} {\sc and} {\sc Rozenberg, G.} 1986.
\newblock Infinitary languages: Basic theory and applications to concurrent
  systems.
\newblock In {\em Current Trends in Concurrency}, {J.~de~Bakker}, {W.-P.
  de~Roever}, {and} {G.~Rozenberg}, Eds. Lecture Notes in Computer Science,
  vol. 224. Springer Verlag, 266--342.

\bibitem[\protect\citeauthoryear{Kelley}{Kelley}{1955}]{kelley55}
{\sc Kelley, J.~L.} 1955.
\newblock {\em General Topology}.
\newblock Van Nostrand, Princeton.

\bibitem[\protect\citeauthoryear{Manna and Pnueli}{Manna and
  Pnueli}{1992}]{mannapnueli92}
{\sc Manna, Z.} {\sc and} {\sc Pnueli, A.} 1992.
\newblock {\em The Temporal Logic of Reactive and Concurrent
  Systems---Specification\/}, first ed.
\newblock Springer Verlag, New York.

\bibitem[\protect\citeauthoryear{Nitsche}{Nitsche}{1994}]{nitsche94a}
{\sc Nitsche, U.} 1994.
\newblock Propositional linear temporal logic and language homomorphisms.
\newblock In {\em Proceedings of the 3rd International Symposium on Logical
  Foundations of Computer Science (LFCS'94)}, {A.~Nerode} {and} {Y.~V.
  Matiyasevich}, Eds. Lecture Notes in Computer Science, vol. 813. Springer
  Verlag, Saint Petersburg, Russia, 265--277.

\bibitem[\protect\citeauthoryear{Nitsche}{Nitsche}{1998a}]{nitsche98}
{\sc Nitsche, U.} 1998a.
\newblock Application of formal verification and behaviour abstraction to the
  service interaction problem in intelligent networks.
\newblock {\em Journal of Systems and Software\/}~{\em 40,\/}~3 (March),
  227--248.

\bibitem[\protect\citeauthoryear{Nitsche}{Nitsche}{1998b}]{nitschediss}
{\sc Nitsche, U.} 1998b.
\newblock {\em Verification of Co-Operating Systems and Behaviour Abstraction}.
  GMD Research Series, vol.~7.
\newblock GMD, Sankt Augustin, Germany.
\newblock Publication of PhD thesis. ISBN: 3-88457-331-4.

\bibitem[\protect\citeauthoryear{Nitsche and Ochsenschl\"ager}{Nitsche and
  Ochsenschl\"ager}{1996}]{nitscheochsenschlaeger96a}
{\sc Nitsche, U.} {\sc and} {\sc Ochsenschl\"ager, P.} 1996.
\newblock Approximately satisfied properties of systems and simple language
  homomorphisms.
\newblock {\em Information Processing Letters\/}~{\em 60}, 201--206.

\bibitem[\protect\citeauthoryear{Nitsche and Wolper}{Nitsche and
  Wolper}{1997}]{nitschewolper97}
{\sc Nitsche, U.} {\sc and} {\sc Wolper, P.} 1997.
\newblock Relative liveness and behavior abstraction (extended abstract).
\newblock In {\em Proceedings of the 16th ACM Symposium on Principles of
  Distributed Computing (PODC'97)}. Santa Barbara, CA, 45--52.

\bibitem[\protect\citeauthoryear{Ochsenschl\"ager}{Ochsenschl\"ager}{1992}]{oc%
hsenschlaeger92}
{\sc Ochsenschl\"ager, P.} 1992.
\newblock {V}erifikation kooperierender {S}ysteme mittels schlichter
  {H}omomorphismen.
\newblock Arbeitspapiere der GMD 688, Gesellschaft f\"ur Mathematik und
  Datenverarbeitung (GMD), Darmstadt. Oktober.

\bibitem[\protect\citeauthoryear{Ochsenschl\"ager}{Ochsenschl\"ager}{1994}]{oc%
hsenschlaeger94a}
{\sc Ochsenschl\"ager, P.} 1994.
\newblock Verification of cooperating systems by simple homomorphisms using the
  product net machine.
\newblock In {\em Workshop: Algorithmen und Werkzeuge f\"ur Petrinetze},
  {J.~Desel}, {A.~Oberweis}, {and} {W.~Reisig}, Eds. Humboldt Universit\"at
  Berlin, 48--53.

\bibitem[\protect\citeauthoryear{Ochsenschl\"ager}{Ochsenschl\"ager}{1995}]{oc%
hsenschlaeger95}
{\sc Ochsenschl\"ager, P.} 1995.
\newblock Compositional verification of cooperating systems using simple
  homomorphisms.
\newblock In {\em Workshop: Algorithmen und Werkzeuge f\"ur Petrinetze},
  {J.~Desel}, {H.~Fleischhack}, {A.~Oberweis}, {and} {M.~Sonnenschein}, Eds.
  Universit\"at Oldenburg, 8--13.

\bibitem[\protect\citeauthoryear{Pnueli}{Pnueli}{1977}]{pnueli77}
{\sc Pnueli, A.} 1977.
\newblock The temporal logic of programs.
\newblock In {\em Proceedings of the 18th Annual IEEE Symposium on Foundations
  of Computer Science}. 46--57.

\bibitem[\protect\citeauthoryear{Thomas}{Thomas}{1990}]{thomas90}
{\sc Thomas, W.} 1990.
\newblock Automata on infinite objects.
\newblock See \citeN{HTCS90B}, 133--191.

\bibitem[\protect\citeauthoryear{Ultes-Nitsche and St~James}{Ultes-Nitsche and
  St~James}{2000}]{ultes-nitschestjames00}
{\sc Ultes-Nitsche, U.} {\sc and} {\sc St~James, S.} 2000.
\newblock Weakly continuation-closed abstractions can be defined on trace
  reductions.
\newblock In {\em Proceedings of the International Workshop on Verification and
  Computational Logic (VCL'2000)}, {M.~Leuschel}, {A.~Podelski},
  {C.~Ramakrishnan}, {and} {U.~Ultes-Nitsche}, Eds. University of Southampton,
  11 pages.

\bibitem[\protect\citeauthoryear{van Leeuwen}{van Leeuwen}{1990}]{HTCS90B}
{\sc van Leeuwen, J.}, Ed. 1990.
\newblock {\em Formal Models and Semantics}. Handbook of Theoretical Computer
  Science, vol.~B. Elsevier.

\bibitem[\protect\citeauthoryear{Vardi and Wolper}{Vardi and
  Wolper}{1994}]{VW94}
{\sc Vardi, M.~Y.} {\sc and} {\sc Wolper, P.} 1994.
\newblock Reasoning about infinite computations.
\newblock {\em Information and Computation\/}~{\em 115,\/}~1 (November), 1--37.

\bibitem[\protect\citeauthoryear{Wolper and Godefroid}{Wolper and
  Godefroid}{1993}]{wolpergodefroid93}
{\sc Wolper, P.} {\sc and} {\sc Godefroid, P.} 1993.
\newblock Partial-order methods for temporal verification.
\newblock In {\em CONCUR'93}, {E.~Best}, Ed. Lecture Notes in Computer Science,
  vol. 715. Springer Verlag, 233--246.

\end{thebibliography}
\bibliographystyle{acmtrans}

\end{document}